\documentclass[prd,aps,twocolumn,nofootinbib,preprintnumbers]{revtex4}
\usepackage{amsmath,amssymb,epsf,epsfig}
\usepackage{graphicx,color}

\def\be{\begin{equation}}
\def\ee{\end{equation}}
\def\bea{\begin{eqnarray}}
\def\eea{\end{eqnarray}}

\def\ltap{\ \raise.3ex\hbox{$<$\kern-.75em\lower1ex\hbox{$\sim$}}\ }
\def\gtap{\ \raise.3ex\hbox{$>$\kern-.75em\lower1ex\hbox{$\sim$}}\ }
\def\gl{\ \raise.5ex\hbox{$>$}\kern-.8em\lower.5ex\hbox{$<$}\ }
\def\roughly#1{\raise.3ex\hbox{$#1$\kern-.75em\lower1ex\hbox{$\sim$}}}

\newcounter{oldcounter}
\addtocounter{equation}{1}
\begin{document}

 \preprint{DESY 11-137}

\title{\bf{Cosmology of Ho\v{r}ava-Lifshitz $f(R)$ Gravity}}
\medskip\
 \author{Sayan K. Chakrabarti}
 \email[Email: ]{sayan.chakrabarti@ist.utl.pt}
 \affiliation{CENTRA, Departamento de F\'{\i}sica, 
 Instituto Superior T\'ecnico, Universidade T\'ecnica de Lisboa - UTL,
 Av.~Rovisco Pais 1, 1049 Lisboa, Portugal.}
 \author{Koushik Dutta}
 \email[Email: ]{kdutta@iiserb.ac.in}
 \affiliation{Deutsches Elektronen-Synchrotron DESY, Theory Group, D-22603 Hamburg, Germany\\ and\\ 
 Indian Institute of Science Education and Research, Bhopal, 462 023, India.}
 \author{Anjan A. Sen}
 \email[Email: ]{anjan.ctp@jmi.ac.in}
 \affiliation{Centre of Theoretical Physics, Jamia Millia Islamia, NewDelhi-110025, India.}

%\date{\today}

\begin{abstract}
We discuss the cosmology of recently proposed Ho\v{r}ava-Lifshitz $f(R)$ gravity. In particular, we derive the modified 
Hubble equation that reduces to the standard HL gravity case in appropriate limit. We show how the bounce solutions in this theory 
are modified due to nonlinear effect of $f(R)$ gravity. We also show that in principle, the Universe in this set-up can show cyclic behavior.  

\end{abstract}

\maketitle

\tighten

\draft

\section{Introduction}

Recently Ho\v{r}ava has proposed a theory of quantum gravity which is power-counting renormalizable in the ultraviolet (UV). This has been achieved through an anisotropic scaling 
between space and time, and therefore it violates Lorentz invariance in the UV \cite{Horava:2008jf}. The infrared limit of the theory reproduces General Relativity 
for a particular choice of a parameter, namely $\lambda = 1.$ The breaking of the Lorentz symmetry is performed by a preferred foliation by three dimensional
spacelike hypersurfaces, which in turn splits the coordinate into space and time. It allows one to write the Einstein Hilbert action of the General Relativity with higher
spatial derivatives of the metric. This improves the UV behaviour of the graviton propagator and makes the theory power-counting renormalizable. Moreover, the 
action has only second order time derivatives preventing the presence of ghosts in the theory. However, the explicit breaking of the general covariance by such foliation of 
the space-time introduces a new scalar degree of freedom which shows some unavoidable pathological behaviours \cite{Cai:2009dx}. Subsequently, there were many more proposals to improve the theory and 
overcome these pathological behaviours \cite{Blas:2009qj}. It is also worth mentioning here, that at the one loop quantum level, the violation of Lorentz invariance has 
dramatic effect \cite{Collins:2004bp, Iengo:2009ix} and in order to have a viable Ho\v{r}ave-Lifshitz (HL) gravity much works need to be done towards understanding the renormalization group flows.

Apart from the abovementioned problems in the HL theory, it still is interesting to look at different aspects of this theory from the perspective of cosmology. Many interesting cosmological implications have been found for the HL gravity  \cite{Lu:2009em}. It was found out that 
Friedmann equations in HL gravity contains some terms which behaves as dark radiation with negative coefficient. Therefore it is possible to find bounce solution 
which has a non-singular behaviour of scale factor as the origin of the Universe \cite{Calcagni:2009ar}. Different UV scaling of HL gravity can also lead to scale-invarint cosmological 
perturbations without any need of inflation \cite{Mukohyama:2009gg}.  Given the fact that HL gravity is  a non-relativistic modified gravity, the idea of cyclic cosmology was first realized in this non-relativistic gravity by Cai and Saridakis \cite{sarikadis}. In this work, only the gradient terms of the gravitational action were modified and the model was free from the strong coupling problem which was extensively discussed in Horava-Lifshitz models\cite{Horava:2008jf}.

On the other hand, many recent observations  indicate that the Universe has probably gone through two phases of accelerated expansion: one in the early epoch
 to solve many cosmological puzzles with the creation of observed density perturbations \cite{Guth:1980zm}, another to accommodate the peculiar observations that the current 
Universe is accelerating \cite{Copeland:2006wr}. Instead of adding some extra matter component in the Universe to invoke the accelerated phases, an alternative is the modification
 of Einstein-Hilbert term by replacing the Ricci scalar $R$ in the action with some general functions  $f(R)$ of the Ricci scalar. For an extensive review see \cite{Nojiri:2008nt}. Considering that HL gravity is  a promising candidate for an UV 
completion of General Relativity, $f(R)$ versions of HL gravity were also proposed \cite{Kluson:2009rk, Kluson:2009xx, Chaichian:2010yi}.

In this paper we make an analysis of the cosmology of $f(R)$ HL gravity that has been proposed in \cite{Kluson:2009rk}. In particular, we will derive the modified 
Hubble equation that reduces to the standard HL gravity case in appropriate limit. Due to nonlinear effect of $f(R)$ gravity we will see how the bounce solutions 
are modified. We will explicitly show that in early time the Universe can go through a bounce phase, whereas in late time it admits turn-around solution. Therefore, in principle, the Universe 
in this set-up can show cyclic behavior. After describing the basics of HL gravity and its cosmology in Sec. 2 and Sec. 3 respectively, we discuss a particular form of 
$f(R)$ HL gravity in Sec 4. In Sec 5 we discuss the cosmology of this particular scenario. 

\section{A Brief Review of Ho\v{r}ava-Lifshitz Gravity}
In this section we briefly review the Ho\v{r}ava-Lifshitz (HL) gravity as was originally proposed in \cite{Horava:2008jf}. The dynamical variables in HL gravity are similar to the ADM formalism of Einstein gravity, namely the lapse function $N$, shift vector $N_i$ and the spatial metric $g_{ij}$. We consider only the projectable version of HL gravity where the lapse function needs to be function of time $t$ only. In terms of these variables the metric is
\bea
ds^2 = -N^2 dt^2 + g_{ij}(dx^{i} + N^{i} dt)(dx^{j} + N^{j} dt),
\eea
where  $N_{i} = g_{ij} N^{j}$. The gravitational part of the action consists of a kinetic and a potential part 
\bea
S_{g} = \int dt d^3 x \sqrt g N \left ( \mathcal{L}_{K} + \mathcal{L}_{V}  \right).
\eea
The kinetic terms are given by 
\bea
\mathcal{L}_{K} = \frac{2}{\kappa^2} (K_{ij} K^{ij} - \lambda K^2),
\eea
where the extrinsic curvature is given by 
\bea
K_{ij}=\frac{1}{2N}(\dot g_{ij}-\nabla_i N_j-\nabla_j N_i),
\eea
with covariant derivatives defined with respect to the spatial metric $g_{ij}$.
The potential term in the ``detailed-balance" \footnote{Without the detailed balance condition, the most general renormalizable theory contains cubic order terms in curvature \cite{Calcagni:2009ar, Kiritsis:2009sh}.} form is given by 
\bea \label{HLaction}
\mathcal{L}_{V} &=&- \frac{\kappa^2}{8}  E^{ij} G_{ij;kl} E^{kl} 
\eea
where the super-metric $G_{ij;kl}$ depends on an arbitrary dimensionless coupling constant $\lambda$ (susceptible to quantum corrections)
\bea
\mathcal{G}_{ij;kl} = \frac{1}{2}(g_{ik} g_{jl} + g_{il} g_{jk}) + \frac{\lambda}{1 - 3 \lambda} g_{ij} g_{kl}.
\eea
The $E$ tensors are given by 
\bea
E^{ij}&=&\frac{2}{w^2}C^{ij}-\mu(R^{ij}-\frac{1}{2}Rg^{ij}+\Lambda_w g^{ij})
\eea
where 
\bea
C^{ij} = \frac{\epsilon^{ikl}}{\sqrt g} \bigtriangledown_{k} (R^j_{~ l} - \frac{1}{4} R \delta^{j}_{~ l})
\eea
is the Cotton tensor, a conserved traceless tensor that vanishes for conformally flat metrics. $\kappa$, $w$, $\mu$ are constants with mass dimension $-1$, $0$, $1$ respectively. $\Lambda_w$ is related to the cosmological constant in the IR limit of the theory. 

Demanding that in the infrared limit, the HL action reduces to the Einstein-Hilbert action
\bea
S_{EH} = \frac{1}{16 \pi G}\int d^4 x \sqrt g N (K_{ij} K^{ij}  - K^2 + R - 2\Lambda)
\eea
allows us to write the speed of light, Newton's constant and the cosmological constant in terms of parameters of the model 
\bea
c = \frac{\kappa^2 \mu}{ 4}\sqrt{\frac{\Lambda_w}{1 - 3 \lambda}}, ~~~ G = \frac{\kappa^2}{32 \pi c}, ~~~  \Lambda = \frac{3}{2} \Lambda_{w}. \label{constants}
\eea  
In addition we have to identify $\lambda = 1$. From Eq.(\ref{constants}) we can easily see that for $\lambda > \frac{1}{3}$, the cosmological constant $\Lambda_w$ must be negative.

\section{Cosmology of HL Gravity}
Now we discuss the implications of HL gravity in the cosmological context \cite{Lu:2009em, Calcagni:2009ar, Kiritsis:2009sh}. In particular, we assume the homogeneous  and isotropic Universe described by the following FRW metric 
\bea
g_{ij}~dx^i dx^j &=& a(t)^2 \left[\frac{dr^2}{1 - k r^2} + r^2 (d\theta^2 + sin^2 \theta  d\phi^2)\right]\nonumber\\
&&
\eea
with $N = 1, ~ N^i = 0$, where  $k =+1 , 0 , -1$ corresponds to a closed, flat or a open Universe respectively. Due to many symmetries of the FRW space-time the calculations simplify by a large amount, but still leaving some novel signatures of the HL gravity. Assuming matter contribution behaves as an ideal fluid we find the following Hubble equation by varying the action with respect to the lapse function $N$ 
\bea
H^2 \equiv \Bigg(\frac{\dot a}{a}\Bigg)^2 &=& \frac{2}{3\lambda - 1} \Bigg(\frac{\Lambda_w}{2} + \frac{\kappa^2}{12} \rho  - \frac{k}{a^2} \nonumber\\ 
&& + \frac{k^2}{2 \Lambda_w a^4} \Bigg) \label{Hubble_HL}.
\eea
Variations with respect to $g_{ij}$ gives us the following equation
\bea
\dot H + \frac{3}{2} H^2 = \frac{1}{3 \lambda -1} \left( \frac{3}{2}\Lambda_w - \frac{\kappa^2}{4}p  - \frac{k}{a^2} - \frac{k^2}{2 \Lambda_w a^4}   \right),
\eea
where $p$ is the pressure of the fluid.

For the vacuum case with $\rho = p = 0$, and we have 
\bea
H^2 = \frac{\Lambda_w}{3\lambda -1} \left(1 - \frac{k}{\Lambda_w a^2} \right)^2.
\eea
As we have noted earlier, for $\lambda > \frac{1}{3}$, the bare cosmological constant $\Lambda_w$ that appears in the HL theory must be negative. Therefore the only solution (static) exists for $k = -1$ with $a^2 = -1/\Lambda_w$. However one can can also make an analytic continuation to extend the cosmological constant into positive regime in the frame of HL gravity. This issue has been already studied explicitly in \cite{Lu:2009em}. In our present investigation, we will only consider the case where the bare cosmologial constant is negative.

We find modifications to the standard Hubble equation due to higher order derivative terms in the action. Interestingly enough, these contributions from higher derivative terms vanish for $k = 0$. In addition, they are important only at small $a$, giving us back to the standard cosmological solutions in the IR. We also see that the Hubble equation is modified by an overall factor depending on $\lambda$. 

The most important modification is the term proportional to $1/a^4$, that contributes a negative energy density for any nonzero spatial curvature. This term is similar to the ``dark radiation'' in the braneworld scenario, with an important difference of being nonzero only for nonzero spatial curvature \cite{Binetruy:1999hy}. If in a collapsing phase (decreasing $a$) of the Universe at a very early time, the matter energy density grows slower than $a^{-4}$, the dynamics leads to a bounce with $H = 0$, and $\dot H > 0$ \cite{Brandenberger:2009yt}. The condition for bounce in HL gravity turns out to be \footnote {Condition is different than what has been found in \cite{Brandenberger:2009yt}.}.
\bea
\left(\frac{\rho}{3} - p \right) > 0.
\eea
 During the contracting phase of the Universe, matter can be described by an oscillating scalar field with a potential
\bea
V(\phi) = \frac{1}{2} m^2 \phi^2.
\eea
During the contraction phase, the  amplitude of oscillation grows until it reaches a critical value where energy density carried by the field becomes constant. Subsequently, due to the rapid contraction of the Universe, negative contributions coming from $1/a^4$ term catches the energy density of $\phi$ field. In the end Universe passes through a bounce when Hubble constant $H = 0$ with $\dot H > 0$. The application of this mechanism to achieve a nonsingular bounce was firstly appeared in the model of Lee Wick theory \cite{lee-wick}.

In the following  section we will find out the cosmological implications of modified HL gravity.  

\section {$f(R)$ Ho\v{r}ava-Lifshitz Gravity}
As an explanation of early and late time acceleration of the Universe, modifications of Einstein gravity without an extra matter source is a viable option (for a general review see \cite{Nojiri:2008nt}). In this approach, the standard Einstein-Hilbert action is added/replaced by the higher order curvature invariants such as $R^2, R_{\mu \nu} R^{\mu \nu}$ etc. Indeed, there are examples of $f(R)$ gravity theories where the early time inflation can be unified with late time cosmic acceleration\cite{nojiri}. On the other hand, Ho\v{r}ava-Lifshitz Gravity is a promising way forward in constructing the theory of quantum gravity \cite{Horava:2008jf}. The theory is power-counting renormalizable in the expense of breaking Lorentz symmetry explicitly. The ultraviolet behavior of the theory is improved by the higher spatial derivative terms of the metric. Subsequently, $f(R)$ HL gravity has been constructed recently following the way the original Einstein gravity was modified to $f(R)$ theories. In this section we review these approaches briefly. 

A straightforward modification of HL gravity has been recently proposed by \cite{Kluson:2009xx, Chaichian:2010yi}
\begin{equation}
S  = \int dt d^3x \sqrt{g} N f(\tilde R), ~~~ 
\end{equation}
with 
\begin{eqnarray}
\tilde R &=&K^{ij} K_{ij} - \lambda K^2 + 2\mu \nabla_{\mu} ( n^{\mu} \nabla_{\nu} n^{\nu} -  n^{\nu} \nabla_{\nu} n^{\mu}) \nonumber\\ &&- E^{ij} \mathcal{G}_{ij;kl}E^{kl},
\end{eqnarray}
where $n^{\mu}$ is a unit vector perpendicular to $t =$ constant space-like hypersurface $\Sigma_t$. Similar to the standard Einstein-Hilbert action, the term proportional to $\mu$ is usually dropped in HL gravity as it is a total derivative term in the action. But in the non-linear generalization to $f(R)$ gravity, this term can not be dropped \cite{Chaichian:2010yi}. Modified version of this HL gravity has been subsequently discussed for unifying inflation with dark energy \cite{Elizalde:2010ep}, and Hamiltonian analysis has been carried out in \cite{Chaichian:2010yi, Kluson:2010xx}. 

Following Ho\v{r}ava's proposal, a different version of modified $f(R)$ HL gravity has been introduced \cite{Kluson:2009rk} early on. In this case, at first a partition function of D-dimensional gravity theory was constructed. Demanding the existence of a $(D+1)$ dimensional quantum gravity theory such that the norm of its ground state wave function coincides with the partition function of D-dimensional theory, it was then shown that infinite number of Hamiltonian obeys the detailed balance conditions. For a special form of the Hamiltonian which corresponds to $f(R) = \sqrt{R}$, a Lagrangian formulation of the theory was also made assuming projectability condition. Action for this modified HL gravity is given by \cite{Kluson:2009rk}

\begin{eqnarray}\label{action}
S = -\kappa^2\int && dtd^3x\sqrt{g^{(3)}}N\Bigg(\sqrt{ {M^6} +\frac{1}{4} E^{ij}\mathcal{G}_{ijkl}E^{kl}} \nonumber\\ 
&\times&\sqrt{ {M^6} -\frac{4}{\kappa^4}(K_{ij}K^{ij}-\lambda K^2)}- {M^6}\Bigg).
\end{eqnarray}
Here $M$ is a mass parameter. In this form of the action, the dimensions of different quantities are as follows: $[G_{ijkl}]= [Mass]^0, [E_{ij}] = [Mass]^3, [\kappa^2]=[Mass]^{-2}$ and $[K_{ij}] = [Mass]^1$. In the original paper by Kluson \cite{Kluson:2009rk}, the action was written in a unit where $M = 1$. In our subsequent calculations, we  also assume $M=1$.

A similar form of the action was also found without projectability condition. In a subsequent paper a generalization for any form of $f(R)$ was presented \cite{Kluson:2009xx}. The new form of $f(R)$ HL gravity are invariant under foliation preserving diffeomorphism similar to the original HL gravity \cite{Horava:2008jf}. We note that the ordinary $f(R)$ gravity theories have full diffeomorphism invariance.
It is trivial to check that linearized version of the above action in Eq. (\ref{action}) gives us back the original HL action. Goal of this article is to present the cosmological consequences of the modified $f(R)$ HL action given by Eq. (\ref{action}).

\section{Cosmology of $f(R)$ Ho\v{r}ava-Lifshitz Gravity}
In this section we will discuss the cosmological solutions of modified $f(R)$ HL gravity defined by Eq. (\ref{action}). In particular, we will look for the `bounce' and `turn-around' solution in this scenario and point out the main differences with the standard HL gravity solutions.

Assuming the background Universe described by a FRW metric and the matter contribution is equivalent to an ideal fluid described by energy density $\rho$ and pressure $p$, the Hubble constant can be extracted as 
\bea
H^2  = && \frac{1}{6(3 \lambda -1)} \left [\kappa^2 \rho - \frac{\rho^2}{2} + 6 \Lambda_w - 12 \frac{k}{a^2} + 6\frac{k^2}{\Lambda_w a^4}  \right]\nonumber\\
&\times& \left(1 - \frac{\rho}{\kappa^2}\right)^{-2}.
\eea
This is the modified Hubble equation for $f(R)$ HL gravity and should be compared with Eq. (\ref{Hubble_HL}). We note that the vacuum solution with energy density $\rho = 0$ is exactly similar to the standard HL gravity theory. Therefore, similar to the standard HL gravity, vacuum solution exists only for open Universe $(k = -1)$ with scale factor $a = \sqrt{-1/\Lambda_w}$, where $\Lambda_w$ is negative for $\lambda > 1/3$. Thus, this particular form of $f(R)$ gravity does not change the pure gravity part in FRW space-time. In contrast to the other version of $f(R)$ HL gravity \cite{Chaichian:2010yi}, the vacuum solution does not admit de-Sitter solution. As we see from Eq. (\ref{action}), similar to the standard HL gravity, many novel terms are proportional to the spatial curvature.  Another important feature of the modified Hubble equation is that even for a flat case ($k=0$), there is a modification in the Einstein equation. This is because of the presence of $\rho^2$ term. Moreover this term comes with a negative sign.

Expanding the above expression for the Hubble constant in terms of $\rho/\kappa^2$ we find
\bea
H^2 &=& \frac{2}{3\lambda - 1} \left(\frac{\Lambda_w}{2} + \frac{\kappa^2}{12} \rho  - \frac{k}{a^2} + \frac{k^2}{2 \Lambda_w a^4}+\frac{\Lambda_w \rho}{\kappa^2} +\frac{\rho^2}{8}  \right. \nonumber \\
 &&\left. +  \frac{3}{2} \frac{\Lambda_w \rho^2}{\kappa^4} - \frac{2k \rho}{a^2 \kappa^2} - \frac{3k\rho^2}{a^2 \kappa^4} + \frac{ k^2 \rho}{a^4 \kappa^2 \Lambda_w} \right.\nonumber\\
&&\left. + \frac{3 \rho^2 k^2}{2 a^4 \kappa^4 \Lambda_w }  \right) + {\mathcal O}(\rho^3)
\eea
Firstly, in the leading order we recover the Hubble equation for standard HL gravity. Higher order terms are due to nonlinear $f(R)$ HL  gravity, and many of them contributes to the ``dark radiations'', $i.e$ redshifts as $a^{-4}$. 

For a flat universe with $k = 0$, from the positivity of $H^2$ we can easily find out that the modified theory must satisfy the following condition
\bea
\kappa^2 - \sqrt{\kappa^2 - 12 |\Lambda_w |} < \rho < \kappa^2 + \sqrt{\kappa^2 - 12 |\Lambda_w |}
\eea 
with $|\Lambda_w | < \kappa^4/12$. Thus the energy density $\rho$ coming from the matter sector is both bounded above and below. This condition is starkly different than the standard HL gravity where we find a lower limit on the energy density $6 |\Lambda_w| < \kappa^2 \rho$ for a given value of $\Lambda_w$. 

Note that the Hubble parameter blows up at the singularity $\rho=\kappa^2$. Although $\dot H$ is also singular at this point, the scale factor, energy density and pressure remains finite. This is a new kind of singularity and can not be classified following \cite{Nojiri:2005sx}. But we should stress that this singularity occurs at the Planck's scale where the classical FRW equation for the cosmological evolution should not be valid and one should consider the quantum gravity effects. So for our purpose, working in the classical regime $\rho< \kappa^2$, avoids hitting this singularity. We comment on the avoidance of this singularity while having bounce solution in the next section.

In the following we will discuss both the early and late time cosmology that is realized with this particular form of $f(R)$ HL gravity. In particular, we will see, whereas in early time the Universe can go through a bounce phase, in late time it admits turn-around solution. In summary, the Universe can go through a cyclic behavior subject to some conditions satisfied by the matter content of the Universe. 

\subsection{Bounce Solution}
One of the important aspect of HL gravity is that it allows the Universe to have bounce solutions without introducing any exotic matter components. As we have discussed earlier, this happens naturally due to ``dark radiation'' term in the Hubble equations. In the modified $f(R)$ HL theory bounce structure changes mainly due to the presence of an extra term in the Hubble equation that depends on the square of energy density and has a negative coefficient. Now, in addition to the standard energy density term, this term also contributes in making the Hubble constant zero as Universe passes through a contraction phase. 

Varying the action with respect to $g_{ij}$ gives us the expression for the time variation of Hubble constant as
\begin{eqnarray}
\dot H &=& \frac{1}{12(3 \lambda -1)} \left[ 3(\rho + p) (\rho - \kappa^2) + \frac{24 k}{a^2} + \frac{24 k^2}{|\Lambda_w| a^4}  \right] \nonumber\\ 
&& \times \left( 1 - \frac{\rho}{\kappa^2} \right)^{-2}-\frac{3(\rho + p)}{\kappa^2} H^2 \left(1 - \frac{\rho}{\kappa^2}\right)^{-1}.
\end{eqnarray}
It is important to remember that $\Lambda_w$ is a negative quantity. A bounce is defined by the condition of $H = 0$, with $\dot H > 0$. Around the bounce the term proportional to $\Lambda_w$ is irrelevant, whereas the term proportional to $a^{-2}$ grows less faster than $a^{-4}$ term, thus we neglect those. As we have noted earlier, the bounce solution is possible only when the Universe is not spatially flat $(k \neq 0)$, and the solution remains same for both open or close Universe. Setting $H = 0$ therefore allows us to find the constraint
\bea
\kappa^2 \rho_{*} - \frac{\rho_{*}^2}{2} - \frac{6 k^2}{|\Lambda_w| a_{*}^4} = 0. \label{bcondn}
\eea 
where $\rho_{*}$ and $a_{*}$ are the energy density and the scale factor respectively when bounce happens.
Now, demanding the positivity of $\dot H$ gives us the bounce condition as 
\bea \label{bounce_condition}
\rho_{*}^2 + \kappa^2 \rho_{*} + 3 p_{*} (\rho_{*} - \kappa^2) > 0.
\eea
If we describe the matter with equation of state parameter $w$ with pressure $p = w \rho$, the above condition can be re-casted conveniently as the following 
\bea \label{bounce_condition1}
\rho_{*} (1 + 3 w) + \kappa^2 (1 - 3 w) > 0.
\eea

Note that (\ref{bcondn}) has two solutions $\rho_*^{\pm} = \kappa^2\pm\sqrt{\kappa^4-\frac{12 k^2}{|\Lambda_{w}|a_{*}^4}}$. If we assume the negative 
sign for the critical density then the bounce occurs at the value of the energy density which is always smaller than the value at the singularity $\rho=\kappa^2$ mentioned in the previous section.

As we can see, a large range of values of $w$ can easily satisfy the bounce condition. The novel feature of having terms in the Hubble equation proportional to $a^{-4}$ saves us from introducing exotic matter for realizing bounce in HL gravity scenario. The most important difference between the condition of Eq.\eqref{bounce_condition1} and the condition for standard HL gravity bounce is the appearance of higher order terms as $\rho^2$. If we model the bounce by a scalar field with nearly flat potential, we have $ w \simeq -1$, and in that case the condition can be written in the simple form 
$\rho_{*} < 2\kappa^2 $. For this case also, we can always avoid the singularity at $\rho=\kappa^2$ mentioned earlier but can have the bounce, if we ensure that the evolution begins after the singularity i.e. $\rho<\kappa^2$. For $w=0$ and $w=1/3$, one can similarly check that the singularity at $\rho=\kappa^2$ can be avoided. As we have seen earlier, for the standard HL gravity case, a bounce is always guaranteed if $ p \simeq - \rho$, but for the modified HL case, there is an upper bound on the total energy density of the bouncing matter field. 

\begin{figure}[ht]
\begin{center}
\mbox{\epsfig{figure=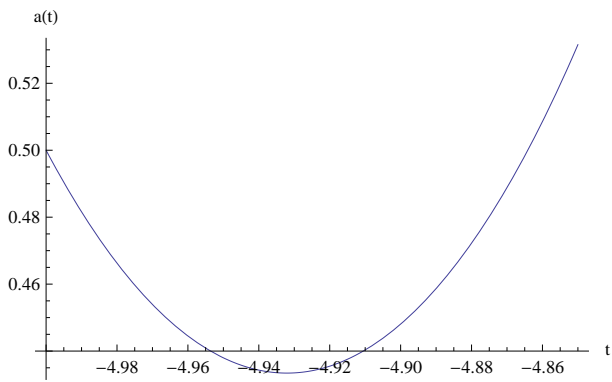, width=8.cm,angle=0}} \caption{{The scale factor $a(t)$ in the bouncing solution (for $w = -1$) shrinks to a non zero value at the bouncing point and 
expands again.}}
\label{plot1}
\end{center}
\end{figure}

\begin{figure}[ht]
\begin{center}
\mbox{\epsfig{figure=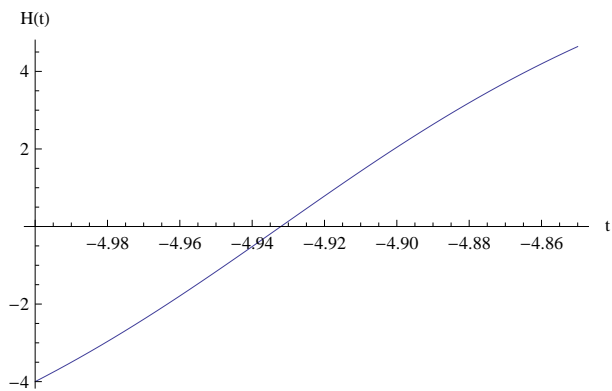,width=8.cm,angle=0}} \caption{{The Hubble parameter $H(t)$ in a bouncing scenario (for $w = -1$) showing that it crosses zero when the universe
goes from a contracting phase to the expanding one.}}
\label{plot2}
\end{center}
\end{figure}

After finding the conditions for the existence of bounce solutions, we now numerically solve the Hubble equation to show the dynamical behavior of a bounce. The 
plots for the scale factor and the Hubble constant are shown in Fig. (\ref{plot1}) and (\ref{plot2}) respectively when energy density is assumed to be a cosmological constant. Without the existence of any exotic matter content, the Universe shrinks to a minimal value of the scale factor and then expands again. At the bouncing point, the Universe evolves through zero Hubble constant.

\subsection{Turnaround Solution}

As we just discussed in the previous subsection, HL gravity naturally admits bounce solution if the Universe starts to contract at some point. In a similar way, the structure of the HL gravity is such that in its expanding cosmological phase, the Universe does not expand forever. For large values of the scale factor the Universe make a turnaround with Hubble constant being zero with $\dot H < 0$. 

Turnaround solution can be easily seen in the standard HL gravity solution. For large value of the scale factor we can easily neglect all terms proportional to the spatial curvature, therefore 
\bea
H^2 = \frac{2}{3 \lambda -1}\left(-\frac{|\Lambda_w|}{2} + \frac{\kappa^2}{12} \rho\right).
\eea
Now it is easy to see that turnaround always happens for non-zero value of cosmological constant with equation of state parameter $w > -1$ at $\rho_* = \frac{6 |\Lambda_w|}{\kappa^2}$. 

In a similar fashion, in the limit of large scale factor, the Hubble constant for the particular $f(R)$ HL gravity that we are considering can be written as
\bea
H^2  =  \frac{1}{6(3 \lambda -1)} \left [\kappa^2 \rho - \frac{\rho^2}{2} + 6 \Lambda_w  \right]
\left(1 - \frac{\rho}{\kappa^2}\right)^{-2}.
\eea
Note that this is valid only for the flat universe with $k=0$. For non flat universe, we can safely ignore $\rho^2$ terms in late times for getting turn around solutions which is similar to the standard HL scenario.
In this case, the condition of $H = 0$ at late time gives us following two solutions for critical turnaround density
\bea
\rho_*^{\pm} = \kappa^2 \left( 1 \pm \sqrt{1 - \frac{12 |\Lambda_w|}{\kappa^2}}   \right).
\eea
The condition for turnaround with $\dot H < 0$ can be written in general as the following
\bea
(1+w)(\rho_* - \kappa^2) < 0
\eea
and can be satisfied with either $w < -1$ and $\rho_* > \kappa^2$, or $w > -1$ and $\rho_* < \kappa^2$. It is clear that the first condition corresponds to the solution $\rho_*^+$, whereas the second condition corresponds to $\rho_*^-$. Firstly, for $\Lambda_w = 0$ the turnaround density becomes $\rho_*^{+} = 2 \kappa^2$, whereas other solution becomes trivial. In this case, the condition for bounce is $w < -1$, i.e the Universe must be dominated by some phantom like field. We note that there is no turnaround solution for standard HL gravity case with $\Lambda_w = 0$. In our case, this feature is due to the presence of negative energy density square term in the Hubble equation. For $\Lambda_w \neq 0$, the relevant solution is $\rho_*^-$ with the condition for bounce being $w > -1$ and $\rho_* < \kappa^2$, and it can be achieved by any matter field satisfying strong  energy condition.

In summary, in this version of modified f(R) HL gravity the Universe can go through alternate cycle of contraction and expansion via bounce and turnaround solution. This is possible even if the cosmological constant $\Lambda_w$ vanishes in our model.

\section{Conclusions}

In conclusion, we have studied the cosmology for the  $f(R)$ gravity model proposed recently by Kluson \cite{Kluson:2009rk}. Our main motivation is to see whether one can get bouncing  as well turnaround solution in this model. There are several interesting features in this model. Firstly  the cosmology in the pure gravity case is exactly similar to what one gets in a standard HL gravity. But in the presence of matter, things change substantially. There is a new correction term proportional to $\rho^2$ which also comes with a negative sign. Due to the presence of such term, the cosmology is different from the  standard one even in the flat case. This is unlike the pure HL gravity, where modification to the standard cosmology only happens for a nonzero spatial curvature. Both the bounce and turnaround conditions in our model  are different from pure HL gravity. Specifically, we need no exotic fluid to get the bounce; one can achieve this with normal matter like radiation.  Similarly turnaround can be achieved even in the absence of a cosmological constant although one may need a phantom fluid for such behaviour.

Overall the $f(R)$ HL gravity has many interesting cosmological signatures as far as the background evolution of the universe is concerned. It will be interesting to see how the inhomogeneities grow in this model and this will be addressed in future works.

\section{Acknowledgement}
S.K.C and K.D would like to thank CTP, Jamia Millia Islamia for a visit when the work was initiated. A.A.S would like to thank MPI, Munich, for a visit where part of the work was done. S. K. C is supported by a grant from FCT (Portuguese Science Foundation). K. D is supported by German Science Foundation (DFG) within the Collaborative Research Center 676 Particles, Strings and the Early Universe. A.A.S acknowledges the financial support from SERC, DST, Govt. of India through the project no.DST-SR/S2/HEP-043/2009 .

\end{document}